\documentstyle[12pt]{article}
\parskip=\medskipamount

\def\appendix#1{
  \addtocounter{section}{1}
  \setcounter{equation}{0}
  \renewcommand{\thesection}{\Alph{section}}
 \section*{Appendix \thesection\protect\indent
 \parbox[t]{11.715cm} {#1}} 
 \addcontentsline{toc}{section}{Appendix \thesection\ \ \ #1}
  }

\renewcommand{\thefootnote}{\fnsymbol{footnote}}

\def \const {{\rm const}}

\newcommand {\cD}{{\cal D}}

\newcommand {\cF}{{\cal F}}
\newcommand {\cG}{{\cal G}}
\newcommand {\cH}{{\cal H}}

\newcommand {\cN}{{\cal N}}

\newcommand {\cW}{{\cal W}}

\newcommand{\bS}{{\bf S}}

\def\a{\alpha}
\def \bi{\bibitem}
\def \ci{\cite}
\def \N {{\cal N}}
\def\b{\beta}

\def\d{\delta}

\def\f{\phi}
\def\g{\gamma}
\def\G{\Gamma}

\def\j{\psi}

\def\l{\lambda}
\def\m{\mu}

\def\o{\omega}
\def\p{\pi}
\def\q{\theta}

\def\s{\sigma}

\def\x{\xi}
\def\z{\zeta}
\def\D{\Delta}
\def\F{\Phi}
\def\J{\Psi}
\def\L{\Lambda}
\def\O{\Omega}

\def\S{\Sigma}
\def\U{\Upsilon}
\def\X{\Xi}
\newcommand{\ad}{{\dot{\alpha}}}                           %new
\newcommand{\bd}{{\dot{\beta}}}                            %new
\newcommand{\ve}{\varepsilon}                            %new
\newcommand{\pa}{\partial}                           %new
\newcommand{\hf}{\frac12}
\newcommand{\1}{\underline{1}}
\newcommand{\2}{\underline{2}}
\newcommand{\sect}[1]{\setcounter{equation}{0}\section{#1}}

\newcommand{\be}{\begin{equation}}
\newcommand{\ee}{\end{equation}}
\newcommand{\bea}{\begin{eqnarray}}
\newcommand{\eea}{\end{eqnarray}}
\newcommand{\non}{\nonumber}

\begin{document}
%%%%%%%%%%%%%%%%%%%%%%%%%
%%%%%%%%%%%%%%%%

\begin{titlepage}
\thispagestyle{empty}

\begin{flushright}
LMU-TPW-99-20\\
OHSTPY-HEP-T-99-027\\
hep-th/9911221 \\
November, 1999
\end{flushright}

%\vspace{5mm}
\begin{center}
{\Large\bf  On  Low-Energy Effective Actions   
in \mbox{$\cN$} = 2, 4 \\
\vspace{3 mm}
Superconformal
 Theories  in Four Dimensions}\\
\end{center}
%\vspace{3mm}

\begin{center} 
{\large I.L. Buchbinder${}^{ {\rm a} ,}$\footnote{Permanent 
address: Department of Theoretical Physics,
Tomsk State Pedagogical University, 
Tomsk 634041, Russia.},
S.M. Kuzenko${}^{\rm b}$, and 
A.A. Tseytlin${}^{{\rm c},}$\footnote{Also at 
Lebedev Physics Institute, Moscow and Imperial College, London}  
}\\
\vspace{2mm}

${}^{\rm a}$\footnotesize{
{\it 
Instituto de F\'{i}sica, Universidade de S\~{a}o Paulo\\
P.O. Box 66318, 05315-970, S\~{a}o Paulo, Brasil}} \\
\vspace{2mm}

${}^{\rm b}$\footnotesize{
{\it Sektion Physik, Universit\"at M\"unchen\\
Theresienstr. 37, D-80333 M\"unchen, Germany} 
} \\
\vspace{2mm}

${}^{\rm c}$\footnotesize{ {\it Department of Physics\\
The Ohio State University  \\
174 W. 18 Avenue, \\
Columbus, OH 43210-1106, USA}}
\end{center}
\vspace{5mm}

\begin{abstract}
\baselineskip=14pt
We study some aspects of  low-energy effective actions 
in 4-d superconformal gauge theories on the Coulomb 
branch. We describe superconformal invariants 
constructed in terms of the
$\cN=2$ abelian vector multiplet  which play the role of  
 building blocks for the $\cN=2,~4$ 
 low-energy effective actions.
We compute the one-loop  effective actions 
in  constant $\cN=2$  field 
strength background 
in $\cN=4$  SYM theory 
and in  $\cN=2$ $SU(2)$ SYM theory  with 
four hypermultiplets in the fundamental representation. 
Using a classification of 
superconformal invariants, we then 
find  the  manifestly $\cN=2$ superconformal 
form  of  these   effective actions.
While our explicit computations are done in the one-loop
approximation, our  conclusions about the 
 structure of the effective actions 
in $\cN=2$ superconformal theories  are  general. 
 We comment on some  relations 
 to supergravity--gauge theory duality 
 in the description of D-brane interactions. 
\end{abstract}
\vfill
\end{titlepage}

\newpage
\setcounter{page}{1}

\renewcommand{\thefootnote}{\arabic{footnote}}
\setcounter{footnote}{0}
%%%%%%%%%%%%%%%%%%%%%%%%%%%
\sect{Introduction}
%%%%%%%%%%%%%%%%%%%%%%%                               
The study of the structure of
 low-energy effective actions in $d=4$ 
superconformal theories  is an  important 
subject from several point of view, 
in particular, in connection with interactions of 
D-branes
in string theory.
Systems of  D3-branes  
have  complementary descriptions in terms of 
gauge theory  and  supergravity.
As one of the consequences, the 
leading-order interaction potential between separated
 branes  admits two equivalent representations:
 as a  classical supergravity potential between a 
  probe and a source, and as a leading 
   term in the  quantum gauge theory
  effective  action. 
  The agreement between the supergravity and the gauge 
  theory expressions for the potential
  is  possible 
  because of the existence of certain non-renormalization
  theorems on the gauge theory side
  (see
  \ci{DS,DG} and references  there). 
  
  One  may conjecture that not only the $F^4/X^4$ term 
   but all higher terms 
\be
\sum^\infty_{n=1} c_n (g^2 N)^{n-1} 
 { F^{2n+2}\over X^{4n}} 
\label{form}
\ee  
in the  Born-Infeld 
   action for a D3-brane 
  probe moving near the  core of a multiple
   D3-brane source
  (or  in  $AdS_5 \times S^5$ space)
  may be reproduced 
  by 
the leading low-energy,   large $N$,  part
of  the { quantum}  $\N=4$  $SU(N)$ SYM effective action.
The latter is 
obtained by keeping  the $U(1)$ $\N=4$  vector multiplet as an
external
 background and
integrating out  massive SYM  fields
(see, e.g., \ci{TA,CT,MM,TR} and refs. there).
This conjecture  seems likely  to be true 
at the first subleading 
order, i.e. for the  $F^6/X^{8}$ term. 
Indeed, it is easy to show
that   this term  is not present 
 in the  $\N=4$   SYM analog  \ci{FT} of the 
 1-loop Schwinger
  effective 
  action, and 
 the result of \ci{BBPT} 
 for the  dimensionally reduced 
  0+1 gauge  theory 
 suggests that  this $F^6$  term should appear in 
  the 2 loop  effective action 
 with precisely the right coefficient 
 to match the supergravity expression. 
 
This conjecture seems, however,   to run into  a problem 
 at   the next order of the $F^8/X^{12}$ term.
According to the supergravity expression
(1.1), it should appear
in the SYM action only at the 3-loop order, but 
 the 1-loop SYM effective action  already 
 contains   $O(F^8)$ term. 
 One 
 may   hope 
that the $F^8$ term does not receive corrections beyond
the 3-loop order,  so that the 3-loop correction dominates
over the  1-loop and 2-loop terms in the supergravity 
limit ($g^2 N \gg 1$).  Still, this may not be enough 
for the agreement since   the $F^8$
invariants in the 1-loop SYM  effective action 
and in the Born-Infeld D3-brane  action 
happen to have different Lorentz index structure.

In order  to shed more light on this  problem 
of the supergravity--SYM  correspondence 
one may   study the constraints 
imposed by the superconformal invariance 
(which is a natural symmetry of the 
supergravity ``D3-brane in $AdS_5 \times S^5$'' action 
\ci{MAL,KAL,MT})
 on the structure of the SYM effective action.\footnote{Some
 implications 
 of special conformal transformations in $\cN = 4$
 SYM theory in the context of AdS/CFT correspondence 
 were considered in \cite{antal}.}
A possible strategy is to start with the 1-loop
 expression for the low-energy effective action on the
 Coulomb branch written in the 
 manifestly superconformally 
 invariant form and try to 
 draw some  general lessons    about the  form  of the 
 effective 
 action which may go beyond the 1-loop order.

 In this paper we 
 shall  consider two superconformal 
theories in four dimensions -- the $\cN=4$ $SU(2)$ SYM
and the $\cN=2$ $SU(2)$ SYM with four hypermultiplets 
in the fundamental representation of $SU(2)$, with 
the gauge group spontaneously broken to its 
$U(1)$ subgroup. 
We  will be  mainly interested in 
the part of their low-energy effective actions 
of $\cN=2,~4$ 
superconformal theories which involves the 
physical bosonic
fields of $\cN=2$ vector multiplet (vector field strength
and scalars).
We  will 
 compute 
the one-loop effective actions in the 
constant field background
\bea
& \cW|_{\q =0} = X = {\rm const}~,\qquad 
 D^i_\a \cW|_{\q =0} = \j^i_\a  = {\rm const}~,& \non \\
& D^i_{( \a} D_{\b)\, i} \cW|_{\q =0} = 8\,F_{\a \b} 
= {\rm const}~, \qquad 
D^{\a (i} D^{j)}_\a \cW |_{\q =0} = 0 ~,&
\label{backg}
\eea
which is a special supersymmetric solution 
of the equations of motion of the abelian $\cN=2$ vector 
multiplet ($\cW$ is the $\N=2$ gauge superfield strength). 
The fact that the  theories
under consideration are superconformal  will allow
 us to 
use 
 the classification of superconformal invariants 
constructed in terms of 
 the abelian $\cN=2$ vector multiplet (section 2). 
 As a result, we will be able 
  to restore 
not only the known $F^4$--type quantum corrections  
\be 
\int {\rm d}^{12} z \, \ \cH(\cW,{\bar \cW})~,\ \ \ \ \ 
\qquad  \cH(\cW,{\bar \cW})\ \propto\ 
\ln {\cW}\, \ln {\bar \cW}
\ee
computed previously  (for $\cN=4$ SYM) 
using  supergraph techniques
\cite{bk1,gr}
(see also \cite{gkpr,bbk,lvu}), but  also all 
terms in  the effective action 
 \bea
\G = c\, \int {\rm d}^{12} z \,  \ 
\ln {\cW}\, \ln {\bar \cW} 
~&+&~ \int {\rm d}^{12} z \,\ \ln {\cW}~ 
\ \L \Big({\bar \cW}^{-2} \, D^4\, 
%D^4\, 
\ln {\cW}\Big)
~+~{\rm c.c.} \non  \\
%&& \qquad
 &+&\  \int {\rm d}^{12} z \, 
\ \U \Big({\bar \cW}^{-2}\,  D^4\, 
%D^4\, 
\ln {\cW} ~ , ~ \cW^{-2}\, 
{\bar D}^4 \, 
%{\bar D}^4\, 
\ln {\bar \cW}
 \Big) ~,
\label{gs}
\eea
which generate the quantum corrections 
of the form (\ref{form}) in components
($\L$ and $\U$ are specific functions
of their arguments).
While  our explicit computations 
will be  done in   the 
one-loop approximation, our 
conclusions about the 
general structure of the effective action 
in superconformal theories
have a universal, loop-independent, character.

The paper is organized as follows. In section 2
we describe superconformal invariants of the
$\cN=2$ abelian vector multiplet, which appear 
as building blocks for the effective actions of
$\cN=2,~4$ superconformal theories.

In section 3 we start with the one-loop effective
action of $\cN=4$ SYM computed for the constant field background
(\ref{backg}), and then restore its  general 
$\cN=2$
 superfield form 
using 
 superconformal invariance 
considerations.
 
 In section 4
the analysis of section 3  
  is extended 
to the case of 
 $\cN=2$ $SU(2)$ SYM with four fundamental 
  hypermultiplets.
We  find that 
a specific  feature of the $\cN=4$ super Yang-Mills 
theory  that there are  no $\L$--type 
 quantum corrections (second term in  (\ref{gs}))
 in the  1-loop effective action
  (in particular, the absence of the  induced $F^6$ 
 term)
 is not shared by generic 
 $\cN=2$ superconformal  theories.
  This unique property  of the $\cN=4$ theory 
   should  be a
   consequence  of a hidden 
$\cN=4$ superconformal symmetry. 

Section 5 contains concluding remarks.
Some useful facts
about $\cN=1,~2$ superconformal transformations
are collected in Appendix.

%%%%%%%%%%%%%%%%%%%%%%
\sect{Superconformal invariants of \mbox{$\cN$} = 2
vector \\ multiplet}
%%%%%%%%%%%%%%%%%%%%%%%%%%%%%

In this section we present superconformal invariants
of an abelian $\cN =2$ vector multiplet described by 
a chiral superfield $\cW(z)$ and its conjugate $\bar \cW (z)$
which are subject to the standard off-shell constraints \cite{gsw}
\bea
&&  {\bar D}_{\ad \,i} \cW = 
%0~, \qquad 
D^i_\a {\bar \cW} = 0~,
\qquad \qquad 
\qquad i = {\1}, {\2} \non \\
&& D^{ij} \cW = {\bar D}^{ij} {\bar \cW}~, 
\qquad \qquad \qquad 
D^{ij} \equiv D^{\a (i} D^{j)}_\a~, \quad 
{\bar D}^{ij} \equiv {\bar D}^{(i}_\ad {\bar D}^{j)\,\ad}~.
\eea 
The $\cN=2$ superconformal transformation law of $\cW$ reads
\be
\d \, \cW = - \x \, \cW - 2 \s \,\cW\ . 
\ee
Here $\x = \x^A D_A$ 
is a superconformal Killing vector, the chiral scalar $\s$ 
is defined by eq. (\ref{lor,weyl}), see Appendix
for more details. It follows then that  
the classical vector multiplet action 
\be 
S_{\rm vm} = \frac{1}{4}\int {\rm d}^4 x \,{\rm d}^4 \q~
\cW^2 
\ee
is, of course,  superconformal invariant.

Let us assume that $\cW$ possesses a non-vanishing expectation
value, as is the case in $\cN$ = 2, 4 superconformal models
with the gauge group  spontaneously broken to its
maximal compact subgroup. Then, using the results of Appendix, 
one check that the following (anti) chiral
combinations\footnote{Here
 $\m$  is  a formal scale which is introduced 
  to make the argument 
 of the 
logarithm  dimensionless. It 
 drops out from all superconformal 
structures listed below.}
\bea
{\bar {\bf \J}}^2 &=& \frac{1}{ 16 {\bar \cW}^2}\, 
D^4\, \ln \frac{\cW}{\m}~, 
\qquad \qquad 
D^4 = (D^{\underline{1}})^2\,
(D^{\underline{2}})^2 \non \\
{\bf \J}^2 &=& \frac{1}{16  \cW^2}\, 
{\bar D}^4\, \ln \frac{\bar \cW}{\m}~, 
\qquad \qquad
{\bar D}^4 = ({\bar D}_{\underline{1}})^2\,
({\bar D}_{\underline{2}})^2 
\eea
transform as scalars with respect to the $\cN=2$ superconformal group,
\be
\d \,{\bar {\bf \J}}^2 = -\x \, {\bar {\bf \J}}^2 ~,
\qquad \d \, {\bf \J}^2 = -\x \, {\bf \J}^2 ~.
\ee
Using the fact that $\cN=2$ superconformal transformations
preserve the $\cN=2$ superspace measure ${\rm d}^{12}z =
{\rm d}^4 x \,{\rm d}^4 \q \,{\rm d}^4 {\bar \q}$,
\be 
(-1)^A\, D_A \, \x^A = 0~,
\ee 
one can construct three types of $\cN= 2$ superconformal 
invariants\footnote{Chiral-like superconformal invariants,
$\int {\rm d}^4 x \,{\rm d}^4 \q\,
\cW^2 H({\bf \J}^2) +{\rm c.c.}$, are equivalent to $\bS_2$.} 
\bea
\bS_1 &=&  \int {\rm d}^{12} z \,
\ln \frac{\cW}{\m}\, \ln \frac{{\bar \cW}}{\m} ~, 
\label{s1} \\
\bS_2 &=& \int {\rm d}^{12} z \,
\L ( {\bar {\bf \J}}^2 )\, \ln \frac{\cW}{\m}   \  
~+~{\rm c.c.} ~, \label{s2} \\
\bS_3 &=& \int {\rm d}^{12} z \, 
\U({\bf \J}^2, {\bar {\bf \J}}^2 ) ~,
\label{s3}
\eea
where  $\L$ and $\U$ are arbitrary holomorphic and real 
analytic 
functions,  respectively\footnote{$\U({\bf \J}^2, {\bar {\bf \J}}^2 )$
is defined modulo K\"ahler--like shifts
$\U({\bf \J}^2, {\bar {\bf \J}}^2 )~ \longrightarrow ~
\U({\bf \J}^2, {\bar {\bf \J}}^2 ) ~+~ \X ({\bf \J}^2)
+ {\bar \X} ({\bar {\bf \J}}^2 )$.}.
These functionals are the main  data 
 describing  quantum corrections of the form
(\ref{form}) (along with special contributions
with derivatives of the 
%scalar and spinor
fields required by supersymmetry)
which  appear in the  low-energy effective actions of 
$\cN$ = 2, 4 superconformal theories.

There exist additional superconformal invariants
constructed  in terms of 
\be
{\bf \S}^{ij} = \frac{1}{ \cW \, {\bar \cW} }\,
D^{ij} \, \cW = \frac{1}{ \cW \, {\bar \cW} }\,
{\bar D}^{ij}\, {\bar \cW}~,
\label{sigma}
\ee
where the primary field $D^{ij}\cW$ transforms as
follows
\be
\d \, D^{ij} \cW = -\x \,   D^{ij} \cW
-2{\rm i}\, \hat{\L}_k{}^{(i}\, D^{j)k} \cW
-2(\s + {\bar \s})\,D^{ij} \cW~.
\ee 
However, ${\bf \S}^{ij}$ involves the free equation 
of motion of the $\cN=2$ vector multiplet. 
As is well-known, contributions to 
%to the one-loop 
effective action, which contain the classical 
equations of motion factor, are ambiguous (in particular, 
gauge dependent).
For that reason  we will ignore 
${\bf \S}$--dependent quantum corrections
in what follows. 

A  large  number of nontrivial 
superconformal invariants can be obtained by 
noting that for a primary superfield $\G_{ij} = \G_{ji}$
with the transformation law 
\be
\d \, \G_{ij} = - \x \, \G_{ij} + 2 \s \, \G_{ij}
+ 2{\rm i}\, \hat{\L}_{(i}{}^k \, \G_{j)\, k} ~,
\label{g-tr}
\ee
its descendant $D^{ij} \,\G_{ij}$ is also primary,
\be 
\d \, D^{ij} \, \G_{ij} = - \x \, D^{ij} \, \G_{ij}
+2 \left( \s - {\bar \s} \right)  D^{ij} \, \G_{ij}~.
\ee
Given an arbitrary function 
$f ({\bf \J}^2, {\bar {\bf \J}}^2 )$,
the superfield $\cW \, f ({\bf \J}^2, {\bar {\bf \J}}^2 )$
transforms like $\cW$, and therefore 
$\G_{ij} \equiv (\cW^2 {\bar \cW} )^{-1} \,
D_{ij} \, \Big( \cW \, 
f ( {\bf \J}^2, {\bar {\bf \J}}^2 ) \Big)$
has  the superconformal transformation law 
(\ref{g-tr}). As a consequence, the following 
combinations 
\bea
&& \cW \, D^{ij}\,\left\{
\frac{1}{\cW^2 {\bar \cW}^2  }\, D_{ij}\, \Big( \cW \, 
f ( {\bf \J}^2, {\bar {\bf \J}}^2 ) \Big) \right\}~, \non \\
&& {\bar \cW} \, {\bar D}^{ij}\,\left\{
\frac{1}{\cW^2 {\bar \cW}^2  }\, D_{ij}\, \Big( \cW \, 
f ( {\bf \J}^2, {\bar {\bf \J}}^2 ) \Big) \right\}
\label{hd1}
\eea
are superconformal scalars. One more possibility
to generate superconformal scalars is to take 
$SU(2)$ invariant products of several superfields
of the form 
\be
\frac{1}{\cW {\bar \cW}  }\, D^{ij}\, \Big( \cW \, 
f ( {\bf \J}^2, {\bar {\bf \J}}^2 ) \Big)
\label{hd2}
\ee
and their conjugates which
transform similar to ${\bf \S}^{ij}$.
Then one can repeat the construction
of superconformal invariants  discussed above 
by replacing the arguments of 
$f ( {\bf \J}^2, {\bar {\bf \J}}^2 )$
by  other superconformal scalars, etc.
%%HERE%%%
 
In this paper, we are mainly interested in 
the part of the low-energy effective action of
  $\cN=2,~4$ 
superconformal theories, which involves the physical bosonic
fields of  $\cN=2$ {\it vector multiplet},
i.e. the $U(1)$ field strength and
 its scalar superpartners,
{ without  higher  derivatives}. 
The crucial point 
is that all relevant component structures
are then  generated by the superconformal invariants 
of the three types  given in 
 (\ref{s1}), (\ref{s2}), (\ref{s3}).
It should be noted that  while 
 many component structures of interest
can be also  obtained from  the 
 superconformal invariants
generated by (\ref{hd1}), (\ref{hd2}) and their descendants,
 the difference between the  %%HERE%%
  two descriptions  is only in 
   terms which involve  higher 
derivatives of the fields.

Let us represent  $\cW$  in terms of its 
$\cN=1$ superfield parts\footnote{Our $\cN=1$
conventions correspond to \cite{BK}.} 
\be 
\cW | = \F~, \qquad \quad D_\a^{\underline{2}}\, \cW |=
2{\rm i}\, W_\a~,
\ee
where we used the notation 
$ U| = U(z)|_{ \q_{ \underline{2} } = 
{\bar \q}^{{\2}} = 0}$, 
for any $\cN = 2$ 
superfield $U$.
{}
Then %%HERE%%
\be
{\bar {\bf \J}}^2 \Big| = 
\frac{1}{4 {\bar \F}^2}\, D^2
\Big( \frac{ W^\a W_\a}{\F^2} + 
\frac{1}{4 \F}\, {\bar D}^2 {\bar \F} \Big)
\equiv {\bar \J}^2  +  
\frac{1}{16 {\bar \F}^2}\, D^2
%\Big(  
{\bar D}^2 \,  \frac{ {\bar \F}  }{\F} 
%{\bar \F} \Big) 
 ~.
\label{projection}
\ee
${}$From the $\cN=1$ superconformal transformations
\bea
\d \, \F &=& - \x \,\F - 2\s \, \F ~, \non \\
\d \, W_\a &=& - \x \,W_\a + {\hat{\o}}_\a{}^\b \, W_\b
- 3 \s \, W_\a~,
\eea
it follows that the (anti) chiral combinations
\be
{\bar \J}^2 = \frac{1}{4 {\bar \F}^2} \,D^2 
\left( \frac{W^2}{\F^2} \right)~, \qquad
\J^2 = \frac{1}{4 \F^2} \,{\bar D}^2 
\left( \frac{{\bar W}^2}{{\bar \F}^2} \right)
\label{jsimple}
\ee
transform as {\it scalars}  with respect to the
$\cN=1$ superconformal group.

%%%%%%%%%%%%%%%%%%%%%%%%%%%%%%%%%%%%%%%%%%%%5
\sect{\mbox{$\cN$} = 4  super Yang-Mills theory}
%%%%%%%%%%%%%%%%%%%%%%%%%%%%%%%%%%%%
In this section we analyze the low-energy effective action 
of the $\cN=4$ $SU(2)$ super Yang-Mills theory with
the gauge group broken to $U(1)$. A generalization to 
the case of an arbitrary semi-simple gauge group
spontaneously broken to its maximal abelian 
subgroup is straightforward and can be done as 
in  refs. \cite{gkpr,bbk,lvu} where 
the leading superfield correction to the low-energy action
was computed.

Our aim will be to find the manifestly superconformal 
invariant
generalization of the well-known Schwinger-type
expression for the  bosonic part of the 
1-loop effective action  of $\N=4$ SYM theory 
in the  purely bosonic
$F_{mn}=\const$ background.
The use of the superconformal 
invariance requirement
may allow, in principle, to go beyond the constant field
approximation. 

${}$For example,  in the $SU(2)$ $\cN=4$ theory  with the 
classical scalar
field  value  producing the mass parameter $X^2= |\Phi|^2$,
the action in the  background $F_{mn} = 
{\rm F}_{mn} {\s_3\over 2} $, \ with ${\rm F}_{mn}$
having eigen-values $f_1$ and $f_2$,  is  given by 
\ci{FT,CTT}\footnote{Here we consider the action in Minkowski space
and hence the sign of $\G$ is opposite to that  in 
\ci{FT,CTT}.}
\begin{equation}
 \Gamma = {4 V_4\over (4\pi)^2}
\int\limits_{0}^{\infty }\frac{{\rm d}t}{t^3} \;{\rm e}^{-t X^2} \,
{ f_1 t \over \sinh f_1 t}\  { f_2 t \over \sinh f_2 t}\
(\cosh  f_1 t - \cosh f_2 t)^2  \ .
\label{yumi}
\end{equation}
Expanding  in powers of $f_n\sim F$ one finds that
there is  no $F^6/X^8$  term, while the $F^8/X^{12}$ 
term has the structure  {\it different}
  from the one that appears in  the 
  expansion of the  abelian BI action (with the scale set up by
  $X$): 
  $$L_{\rm BI} = 
   X^{4} [\sqrt { (1 + f^2_1/X^4) (1 + f^2_2/X^4)} -1 ]
 \ .   $$
The  $F^8$ terms in the BI and SYM actions are thus 
different combinations 
of the $F^8$-type superinvarinats.

Below we shall find how to 
``supersymmetrize'' the bosonic expression (\ref{yumi}).
Using the background field formulation \cite{bbko}
for general $\cN =2$ super Yang-Mills theories
in $\cN=2$ harmonic superspace \cite{gikos},
it was shown \cite{bk1} that under some
restrictions on the background $\cN=2$ vector multiplet
$\cW = \{W_\a \, ,\, \F \}$, the one-loop effective action 
of $\cN=4$ SYM admits a simple functional 
representation in terms of $\cN=1$ superfields
\be
\exp ({\rm i}\, \G) = \int {\rm D} {\bar V}\,
{\rm D} V\, \exp \left\{ {\rm i} \,
\int {\rm d}^8z \, {\bar V} \D V\right\}\ , 
\label{p-i}
\ee
where the operator $\D$ is defined by 
\be
\D = \cD^a \cD_a + W^\a \, \cD_\a 
- {\bar W}_\ad  \, {\bar \cD}^{\ad} - |\F|^2~.
\ee
The integration in (\ref{p-i}) is carried out over 
complex {\it unconstrained} $\cN=1$ superfields $V, {\bar V}$.
The algebra of $\cN=1$ gauge-covariant derivatives is
\bea
& \{ \cD_\a \, ,\, {\bar \cD}_\ad \}  = - 2{\rm i}\, \cD_{\a \ad}~,
\qquad [ \cD_a \, ,\, \cD_b  ] = {\rm i}\, F_{ab} \non \\
& {}[ \cD_{\a \ad} \, ,\, \cD_\b ] = 
-2{\rm i} \, \ve_{\a \b} \,{\bar W}_\ad ~,\qquad
{}[ \cD_{\a \ad}\, , \, {\bar \cD}_\bd  ] = 
-2{ \rm i}\, \ve_{\ad \bd}\, W_\a ~,
\eea
where
\be
F_{\a \ad, \b \bd} = (\s^a)_{\a \ad} (\s^b)_{\b \bd} \,F_{ab}
= -{\rm i} \ve_{\a \b} 
{\bar D}_{(\ad}{\bar W}_{\bd )}
%{\bar F}_{\ad \bd} 
-{\rm i} \ve_{\ad \bd} D_{(\a} W_{\b)} ~.
\ee

${}$For a simple superfield background 
\be
D_\a W_\b = D_{(\a} W_{\b)} = {\rm const}~, \qquad 
\F = {\rm const}
\label{approx}
\ee
the effective action can be exactly computed using
the superfield proper time technique (see \cite{BK} for
a review), and the result is \cite{opb}
\bea
\Gamma&=&
\frac{1}{64\pi^2} \int {\rm d}^8 z \int_0^{\infty}\frac{{\rm d}t}{t^3}~
W^2\bar{W}^2 ~\exp(-t|\Phi|^2) \non \\
& \times &
{\rm tr} \Big[\Big(\frac{{\rm e}^{tM}- {\bf 1}}{M}\Big)
\Big(\frac{{\rm e}^{-tM}- {\bf 1}}{M}\Big)\Big]
~{\rm tr} \Big[\Big(\frac{{\rm e}^{t\bar{M}}-{\bf 1}}{\bar{M}}\Big)
\Big(\frac{{\rm e}^{-t\bar{M}}-{\bf 1}}{\bar{M}}\Big)\Big] ~
{\rm det} \Big(\frac{ tF}{\sin(t F)}
\Big)^{\hf} 
\non \\
&=&
\frac{1}{16\pi^2} \int {\rm d}^8 z \int_0^{\infty}\frac{{\rm d}t}{t^3}~
W^2\bar{W}^2 ~\exp(-t|\Phi|^2) \non \\
& \times &
\det \Big(\frac{{\rm e}^{tM}- {\bf 1}}{M}\Big)
~\det \Big(\frac{{\rm e}^{t\bar{M}}-{\bf 1}}{\bar{M}}\Big)
~{\rm det} \Big(\frac{ tF}{\sin(t F)}
\Big)^{\hf}~,
\label{res}
\eea
where 
\bea
& M_{\alpha}{}^{\beta}=D_{\alpha}W^{\beta} =2{\rm i} F_\a{}^\b~,
\qquad
\bar{M}_{\dot{\alpha}}{}^{\dot{\beta}}= -
\bar{D}_{\dot{\alpha}}\bar{W}^{\dot{\beta}}=
-2{\rm i} {\bar F}_\ad{}^\bd ~. 
\eea
The effective action is  ultraviolet 
and infrared finite.

To bring eq. (\ref{res}) to a more useful form, 
we first note
\be
 {\rm tr} \Big[\Big(\frac{{\rm e}^{tM}- {\bf 1}}{M}\Big)
\Big(\frac{{\rm e}^{-tM}- {\bf 1}}{M}\Big)\Big]
= \frac{4}{B^2}\,\Big(1 - \cosh (tB) \Big) ~,
\ee
where
\be
B^2 \equiv \hf \,{\rm tr} (M^2) = 
\frac{1}{4}\, D^2 W^2~.
\ee
In terms of  the two invariants of the electromagnetic field
\be
\cF = \frac{1}{4} \,F^{ab}\,F_{ab}~, \qquad \qquad
\cG = \frac{1}{4}\,{}^*F^{ab} \,F_{ab}~,
\ee
we find  that for the background  under consideration one
has 
\be
B^2 = 2(\cF + {\rm i}\,\cG)~, 
\ee
and 
\be
\frac{1}{16}\, D^2 {\bar D}^2\,(W^2 {\bar W}^2)
= \frac{1}{16}\, D^2 W^2~{\bar D}^2{\bar W}^2
=B^2 {\bar B}^2 = 4 (\cF^2 + \cG^2)~.
\ee
Then \cite{schwinger}
\bea
{\rm det} \Big(\frac{ tF}{\sin(t F)}
\Big)^{\hf} &=& \frac{ 2{\rm i}\, t^2 \cG }
{ \cosh t\sqrt{ 2(\cF + {\rm i}\,\cG) }
- \cosh t\sqrt{ 2(\cF - {\rm i}\,\cG) } } \non \\
&=& \hf\,\frac{ t^2 (B^2 - {\bar B}^2)  }
{ \cosh(tB ) 
- \cosh (t{\bar B})  }~, 
\eea
and therefore the component form of 
the effective action is 
(which is equivalent to the one in  (3.1))
\bea
\Gamma &=&
\frac{1}{4\pi^2} \int {\rm d}^4 x \int_0^{\infty}\frac{{\rm d}t}{t^3}~
 ~\exp(-t|\Phi|^2) \non \\
& \times & 
\Big( \cosh t\sqrt{ 2(\cF + {\rm i}\,\cG) } -1 \Big)
\Big( \cosh t\sqrt{ 2(\cF - {\rm i}\,\cG) } -1 \Big)\non \\
& \times &
\frac{ 2{\rm i}\, t^2 \cG }
{ \cosh t\sqrt{ 2(\cF + {\rm i}\,\cG) }
- \cosh t\sqrt{ 2(\cF - {\rm i}\,\cG) } } ~.
\eea

The superfield effective action is 
\bea
\Gamma  &=&
\frac{1}{8\pi^2} \int {\rm d}^8 z \int_0^{\infty}
{\rm d}t \,t~
W^2\bar{W}^2 ~\exp(-t|\Phi|^2) \non \\
& \times & \frac{ \cosh (tB) -1}{t^2 B^2} ~
\frac{ \cosh (t\bar B) -1}{t^2 {\bar B}^2}~
 \frac{ t^2(B^2 - {\bar B}^2 )}
{ \cosh (tB) - \cosh (t\bar B) }~.
\eea
After a simple rescaling of the proper-time integral, we 
can rewrite the action as follows
\bea
\Gamma &=&
\frac{1}{8 \pi^2} \int {\rm d}^8 z \int_0^{\infty}
{\rm d}t \,t\, {\rm e}^{-t} ~
\frac{ W^2\bar{W}^2}{\F^2 {\bar \F}^2}  \non \\
& \times & \frac{ \cosh (t \J) -1}{t^2 \J^2} ~
\frac{ \cosh (t\bar \J) -1}{t^2 {\bar \J}^2}~
 \frac{ t^2(\J^2 - {\bar \J}^2 ) }
{ \cosh (t \J ) - \cosh (t \bar \J) }~,
\label{fin}
\eea
with $\J$ and $\bar \J$ defined in eq. 
(\ref{jsimple}).

Let us introduce the following function
\bea 
&& \o (x, y) =\o (y, x) = \frac{ \cosh x - 1}{x^2}\,
\frac{ \cosh y - 1}{y^2}\,
\frac{x^2 - y^2}{ \cosh x -  \cosh y} -\hf \non \\
&&\o (0 , y) = \o (x, 0) = 0~.
\label{varphi}
\eea
Then the effective action can be rewritten in the form
\bea
\Gamma &=&
\frac{1}{16 \pi^2} \int {\rm d}^8 z
\frac{ W^2\bar{W}^2}{\F^2 {\bar \F}^2} \non \\
&+& 
\frac{1}{8 \pi^2} 
\int {\rm d}^8 z 
\int_0^{\infty}
{\rm d}t \,t\, {\rm e}^{-t}~
%\,\frac{t}{\exp (t) }~
% \int {\rm d}^8 z ~
\frac{ W^2\bar{W}^2}{\F^2 {\bar \F}^2}  \,
\o(t \J\, ,\, t \bar \J)~.
\label{fin2}
\eea

Now we come to the { key point}.
Untill now  we have used the constant field approximation
(\ref{approx}). However, in eq. (\ref{fin2}) we 
may no longer
assume such an approximation. The effective 
action of $\cN=4$ SYM
should be superconformal
invariant,  but $\J$ and $\bar \J$ are basically 
the only superconformal scalars constructed from 
both $W_\a$ and $\F$ (modulo contributions
involving the free equations of motion terms
$ D^\a W_\a$ and $D^2 \F$ and higher derivative invariants,
see sec. 2). 
Thus the  effective action (\ref{fin2}) is manifestly
invariant under 
$\cN=1$ superconformal transformations !

Of course, the effective action should  not only be 
manifestly $\cN=1$ superconformal, but
 $\cN=2$ superconformal as well.
One can restore a $\cN =2$ superconformal form 
of $\G$ simply by noting that $ \J$
 is a part (\ref{projection}) 
of the leading $\cN=1$ component
of $ {\bf \J}$.

As  follows from (\ref{varphi}) and (\ref{fin2}),
$\G$  contains
contributions of the  two types
\bea
S_1 &=& \int {\rm d}^8 z \,
\frac{ W^2\bar{W}^2}{\F^2 {\bar \F}^2}~, \\
S_3^{(m,n)} &=&
\int {\rm d}^8 z \,
\frac{ W^2\bar{W}^2}{\F^2 {\bar \F}^2}
\J^{2m}\,{\bar \J}^{2n}~, \qquad \quad m,n \neq 0~.
\eea
Using the identities
\bea 
\frac{1}{16} D^4\, \ln \cW \Big| &=& 
\frac{1}{4}D^2 (\frac{W^\a W_\a}{\F^2}) ~+~ \dots~, \non \\
\frac{1}{4} (D^{\underline{2}})^2 
\frac{1}{\cW^{2m}} \Big| &=& 
%-2m(2m+1) \frac{W^\a W_\a}{\F^{2m+2}}  ~+~ \dots 
%\non \\ = 
-\frac{ 2m(2m+1) }{\F^{2m}}\, \frac{W^\a W_\a}{\F^2}
 ~+~ \dots ~, 
\eea 
where dots denote terms involving derivatives
of $\F$, we observe that the  $\cN=2$ extensions of
$S_1$ and $S_3$ are
\bea
\bS_1 &=&  \int {\rm d}^{12} z \,
\ln \frac{\cW}{\m}\, \ln \frac{{\bar \cW}}{\m} ~, \\
\bS_3^{(m,n)} &=& \frac{1}{2m(2m+1)2n(2n+1)}\,
 \int {\rm d}^{12} z \,
{\bf \J}^{2m}\,{\bar {\bf \J}}^{2n}~.
\eea
Let $\O (x,y)= \O (y,x)$ be the analytic function 
related to $\o (x,y)$ as follows: if 
\be 
\o(x,y) = \sum_{m,n =1}^{\infty} 
c_{m,n}\, x^{2m} \,y^{2n}~, 
%\qquad
\ee
then 
\be
\O(x,y) = \frac{1}{4}\sum_{m,n =1}^{\infty}
\frac{ c_{m,n} }{ m(2m+1)n(2n+1) }
\, x^{2m} \,y^{2n}~.
\ee
Then the manifestly $\cN=2$ 
superconformal form of $\G$
is
\bea
\Gamma &=&
\frac{1}{16 \pi^2} \int {\rm d}^{12} z \,
\ln \frac{\cW}{\m}\, \ln \frac{{\bar \cW}}{\m} ~,\non \\
&+& 
\frac{1}{8 \pi^2}  \int {\rm d}^{12} z
\int_0^{\infty}
{\rm d}t \,t\, {\rm e}^{-t}~
%\,\frac{t}{\exp (t) }~
\O(t {\bf \J}\, , \, t \bar {\bf \J}) ~.
\label{fin3}
\eea 
Here the first term was computed in \cite{bk1,gr}
(see also \cite{gkpr,bbk,lvu}).

As is seen from (\ref{fin3}), the one-loop effective 
action of $\cN=4$ SYM does not contain terms 
  described by the ``second'' superconformal  invariant
  (\ref{s2}).
In particular,  there are  no $F^6$-type 
corrections
generated by 
\be
\int {\rm d}^{12} z \,\frac{1}{{\bar \cW}^2}\,
\ln \frac{\cW}{\m}\, D^4 \ln \frac{\cW}{\m}~.
\ee
Such terms are expected to appear at  the 2-loop 
order.

The  absence of this ``$F^6$'' correction at the 
1-loop order 
  is a unique  feature of the  maximally supersymmetric 
$\cN=4$ super Yang-Mills theory
(which,  as discussed in the Introduction,
is crucial for supergravity--SYM correspondence
at the subleading order).
As we are going 
to demonstrate in the next section, 
this property  is no longer true 
in  generic  $\cN=2$ superconformal models.

It  may be  instructive to compare the low-energy action
(\ref{fin2}) with the $\cN=1$ supersymmetric Born-Infeld
action \cite{bi} 
%(see \cite{TR} for a review) 
\bea
S_{\rm BI} &=& 
 \frac{1}{4}\int {\rm d}^6z \, W^2 +
\frac{1}{4}\int {\rm d}^6{\bar z} \,{\bar  W}^2 
+ { 1 \over X^4} \,  \int {\rm d}^8z \, \frac{W^2\,{\bar W}^2  }
{ 1 + \hf\, a \, + 
\sqrt{1 + a +\frac{1}{4} \,b^2} }~,
\label{bi} \\
 a &=&   { 1 \over 2X^4} \, 
\Big(D^2\,W^2 + {\bar D}^2\, {\bar W}^2 \Big)~,
\qquad
b= \, { 1 \over 2X^4} \, \Big(D^2\,W^2 - 
{\bar D}^2\, {\bar W}^2 \Big)~, \non 
\eea
where we  used $1/X$ as a scale parameter. 
The non-trivial last term here has the structure similar to 
that of  $\Gamma $ in (\ref{fin2}), with 
$X^2$ playing the role of $|\F |^{2}$.
While the two actions coincide at the 
leading $W^2 \bar W^2$ order, they 
contain different combinations of 
invariants at higher orders (see also the discussion in
Introduction).
In particular, the subleading ``$F^6$" term  
which was absent in the  1-loop  $\cN=4$ SYM effective 
action  is 
present in the BI action (\ref{bi}) and has the form 
\be
- \frac{1}{8X^8} \,  \int {\rm d}^8z \,W^2\,{\bar W}^2
\Big(D^2\,W^2 + {\bar D}^2\, {\bar W}^2 \Big) ~.
\ee

%%%%%%%%%%%%%%%%%%%%%%%%%%%%%%%%%%%%%%%%%%%%%%%%%%%
\sect{\mbox{$\cN$} = 2 superconformal models}
%%%%%%%%%%%%%%%%%%%%%%%%%%%%%%%%%%%%%%%%%%%%%%%%%

In this section  we shall
consider a special $\cN=2$ superconformal 
theory --  the $\cN=2$  
$SU(N)$ super Yang-Mills model with $2N$
hypermultiplets in the fundamental representation;
the effective action of generic  $\cN=2$ 
superconformal models
\cite{hsw} can be analyzed in a similar fashion.
For simplicity, only the case of 
 $N = 2$
will be discussed, with the gauge group $SU(2)$
spontaneously broken to its $U(1)$ subgroup.

Both $\cN=2$  SYM  
and hypermultiplet models are
superconformal invariant at the classical level. 
Their quantum effective
actions
include the  scale independent 
non-holomorphic terms besides
standard
divergent and holomorphic scale dependent contributions. 
For special combinations of these models
 divergent
 and holomorphic
contributions 
cancel out and the full quantum effective 
action is  superconformal invariant.

${}$For computing the one-loop low-energy effective
action
of a hypermultiplet coupled to a background abelian $\cN=2$
vector multiplet
it is sufficient to make use of the simplest 
realization of the   hypermultiplet
in terms of two $\cN=1$ covariantly chiral 
superfields 
$\f_1$ and $\f_1$  with  opposite $U(1)$ charges
$e = \pm 1$, 
%${\bar \cD}_\ad  \f_{1,2} = 0$, 
with the action
\be
S =  \int {\rm d}^8 z \, \left( 
{\bar \f_1 } \f_1 +
{\bar \f_2 } \f_2 \right)
~+~  \left\{ \,
 {\rm i} \, \int {\rm d}^6 z\, \F \, 
\f_1 \, \f_2
~+~{\rm c.c.} \right\}~,\qquad  {\bar \cD}_\ad  \f_{1,2} = 0~. 
\ee
In the constant field approximation (\ref{approx}), 
the effective action is given by a functional determinant
of the D'Alambertian
\be
\D_{\rm c} = \cD^a \cD_a + W^\a \cD_\a - |\F|^2 \ , 
\ee
which acts on the space of covariantly chiral superfields.
The effective action is \cite{sqed,mag,opb}
\bea
\G_{\rm hm} &=&
\frac{1}{16\pi^2}  \int_{\epsilon^{2}}^{\infty}
\frac{ {\rm d}t}{t}~\exp(-t|\Phi|^2)~
 \int {\rm d}^6 z\,
W^2  \non \\
& \times & \frac{ \cosh (tB) -1}{t^2 B^2} ~
\frac{ (B^2 - {\bar B}^2 )t^2 }
{ \cosh (tB) - \cosh (t\bar B) }~,
\label{h-mea}
\eea
where  $\epsilon \to 0 $ is a  UV cutoff.

The form of $\G_{\rm hm}$  
is determined by the function
\be 
\l (x, y) = \frac{ \cosh x - 1}{x^2}\,
\frac{x^2 - y^2}{ \cosh x -  \cosh y}~,
\qquad  \quad \l (x, 0) = 1~.
\ee
%with the following simple property
%\be
%\l (x, 0) = 1~.
%\ee 
It is useful to introduce a new function
$\z (x,y)$  related to $\l$ by 
\bea
\l (x,y) -1 &=& - y^2 \z (x,y)~, \non \\
\z (x,y)= \z (y,x) &=&
 \frac{ y^2 ( \cosh x - 1) -
x^2 ( \cosh y - 1) }
{x^2 y^2 (\cosh x -  \cosh y )}~.
%\non \\
%\z (x, 0) &=& \frac{ \cosh x - 1 - \hf x^2}
%{x^2(\cosh x - 1)} ~, \non \\
%\z (0,0) &=& = \frac{1}{12}~.
\eea 
Recalling the definition 
${\bar B}^2 = \frac{1}{4} {\bar D}^2 {\bar W}^2$,
we can rewrite the effective action 
 as follows
\bea
\G_{\rm hm} &=&
\frac{1}{16\pi^2}  \int_{ \epsilon^2 }^{\infty}
\frac{ {\rm d}t}{t}~\exp(-t|\Phi|^2)~
 \int {\rm d}^6 z\,
W^2  \non \\
&  &- \frac{1}{16\pi^2}  \int_0^{\infty}
\frac{ {\rm d}t}{t}~\exp(-t|\Phi|^2)~
 \int {\rm d}^6 z\,
W^2  \,t^2 \,{\bar B}^2 ~\z (tB, t{\bar B}) \ , 
\eea
i.e. 
\bea
\G_{\rm hm} 
&=& - \frac{1}{16\pi^2}  
 \int {\rm d}^6 z\,
W^2 
\ln \frac{\F}{\m} ~+ ~{\rm c.c.}
 \non \\
& &+  \frac{1}{16\pi^2}  
\int {\rm d}^8 z\,
\int_0^{\infty}
 {\rm d}t\,t\, {\rm e}^{-t}~
%\frac{t}{\exp(t)}~
% \int {\rm d}^8 z\,
\frac{W^2 {\bar W}^2}{\F^2 {\bar \F}^2} 
~\z (t {\bar \J}, t \J)\ ,
\label{h-mea2}
\eea
where we have absorbed  the UV cutoff
 into  the
renormalization scale $\m$.
Here the  first term (holomorphic contribution) 
%\be
%- \frac{1}{16\pi^2}   \int {\rm d}^6 z\, W^2 
%\ln \frac{\F}{\m} ~+~{\rm c.c.} \ee
may be  derived also  by other  well known 
methods\footnote{In obtaining eq. (\ref{h-mea2}), 
we concentrated on the quantum corrections involving 
the vector multiplet strength and did not take into 
account the effective K\"ahler potential 
$K(\F, {\bar \F}) = - \frac{1}{16 \p^2}{\bar \F} \F\, 
\ln ({\bar \F} \F /\m^2) ={\bar \F}\, \cF{}'(\F) + 
\F \,{\bar \cF}'( {\bar \F}) $ 
generated by the holomorphic Seiberg potential 
$\cF (\F) = - \frac{1}{32 \p^2}\F^2 
\ln (\F/ \m)$. A derivation of $K(\F, {\bar \F})$
in the framework of the superfield
proper time technique, which we used in this 
paper, can be found in \cite{bky,BK}.}
(see, e.g. \cite{bbiko}).

In the  $\cN=2$ superconformal theories  
 holomorphic contributions cancel out. 
Let us recall how this  happens  for the present 
model with 
 4  fundamental hypermultiplets.
 Each hypermultiplet has two $SU(2)$ components, so that 
altogether we have 8 abelian 
hypermultiplets with charges $e= \pm \hf$
with respect to the unbroken $U(1)$ 
generated by $\hf \,\s_3 $. 
In addition, we have 
the adjoint ghost superfields or  two 
hypermultiplets with 
$U(1)$ charges $ e= \pm 1$. 
The charges may be  accounted for 
by replacing $W_\a$ and $\F$ in the effective action  by 
\be
W_\a ~\rightarrow ~ e W_\a~, \qquad
\F ~\rightarrow ~ e \F~.
\ee
Then the complete effective action is
\bea
\G &=& 8 \times \frac{1}{16\pi^2} \int {\rm d}^8 z\, 
\int_0^{\infty}
 {\rm d}t \,t\, {\rm e}^{-t}~
\frac{W^2 {\bar W}^2}{\F^2 {\bar \F}^2} 
~\z ( 2 t  \J , 2 t  {\bar \J}) \non \\
& -&  2 \times \frac{1}{16\pi^2} \int {\rm d}^8 z\, 
\int_0^{\infty}
 {\rm d} t \,t\, {\rm e}^{-t}~
\frac{W^2 {\bar W}^2}{\F^2 {\bar \F}^2} 
~\z ( t \J, 
 t {\bar \J}) \non \\
&+ & \frac{1}{8\pi^2} \int {\rm d}^8 z\, 
\int_0^{\infty}
 {\rm d}t \,t\, {\rm e}^{-t}~
\frac{W^2 {\bar W}^2}{\F^2 {\bar \F}^2} 
\Big\{\o ( t \J, 
 t  {\bar \J}) + \hf \Big\}~,
\label{last}
\eea
with the function $\o (x,y)$ defined
in (\ref{varphi}). 
Here the last term
coincides with the effective action (\ref{fin2})
of $\cN=4$ SYM\footnote{The $\cN=4$ SYM theory 
is equivalent to the $\cN=2$ SYM 
coupled to a single hypermultiplet in the adjoint representation; 
in this case, the hypermultiplet and the ghost contributions 
cancel each other, and the effective action is given by the last
term in (\ref{last}).}.  

Note that since 
\be
\z (x, 0) = \frac{ \cosh x - 1 - \hf x^2}
{x^2(\cosh x - 1)} ~,
\ee
 the effective action now contains the $\cN=2$
superconformal invariants 
of the type (\ref{s2}) (and, in particular, the ``$F^6$" 
contributions (3.28)) which were 
absent in the $\cN=4$ case.
%\vspace{5mm}

%\noindent
%%%%%%%%%%%%%%%%%%%%%%%%%%%%%%%%%%%%%%%%%%%%%%%%%%%
\sect{Conclusions}
%%%%
Let us  summarize the results obtained.

We described the superconformal invariants which are
constructed in terms of the $\cN=2$ abelian vector
multiplet and play the role of building blocks
for the low-energy effective actions
of $\cN=2$ or $\cN=$4 superconformal theories on the 
Coulomb branch.
We then computed the one-loop effective actions 
in  constant $\cN=2$  field strength background 
in $\cN=4$ SYM theory 
and in a particular  $\cN=2$ $SU(2)$ gauge  theory.
% with four hypermultiplets in the fundamental 
%representation.  

The fact that the  theories
under consideration are {\it superconformal}, allowed
us to go beyond the constant field approximation 
and to restore, with the aid of 
the classification of superconformal invariants,
the one-loop effective actions (\ref{gs})
(with $\L$ and $\U$ being special model-dependent 
functions). These actions generate
contributions which in components  have the  form 
  (\ref{form})
 (with no coupling constant prefactors since we 
 consider the 1-loop approximation).

The crucial difference between the $\cN=4$ SYM theory and
generic $\cN=2$ superconformal models is that 
the second term in (\ref{gs}) is absent at the one-loop 
level in $\cN=4$ SYM. The first term in (\ref{gs}),
which generates $F^4$--corrections, is known to 
be one-loop exact \cite{DS}. 
It would be of interest to study if there are  
possible non-renormalization 
theorems for the quantum corrections
which are given by the second and the third terms in
(\ref{gs}) for particular  choices of the function $\L$
and $\U$.
% (see \cite{DG} for a recent discussion 
%of   issues). 

\vskip.5cm
%%%%%%%%%%%%%%%%%%%%%%%%%%%%%%%%%%%%%%%%%%%%%%%%%%%%
\setcounter{section}{0}
\setcounter{subsection}{0}

\noindent
{\bf Acknowledgements} \hfill\break
I.B. and S.K. are glad to acknowledge the participation 
of A. Petrov at early stage of this work.
S.K. benefited from discussions with E. Ivanov 
and S. Theisen.
We would like to acknowledge 
the support of the NATO collaborative research grant
 PST.CLG 974965.
The work of I.B. and S.K. was partially supported
by the RFBR grant No.99-02-16617, by the INTAS grant
No.96-0308 and by the 
DFG-RFBR grant No.99-02-04022. 
I.B. was also supported by the 
GRACENAS grant No.97-6.2-34,  FAPESP  and thanks 
the Institute of Physics at the 
University of Sao Paulo for hospitality.
The work of S.K. was supported in part by the 
`Deutsche Forschungsgemeinschaft'.
The work of A.T. was  supported in part by
the  DOE grant  DOE/ER/01545-779, 
by the EC TMR   grant ERBFMRX-CT96-0045 and  by the 
INTAS grant No.96-538.

%%%%%%%%%%%%%%%%%%%%%%%%%%%%%%%%%%%%%%%%%%%%%%%%%%%%
\setcounter{section}{0}
\setcounter{subsection}{0}

\appendix{
 Superconformal transformations}

%\sect{\mbox{$\cN$} = 1, 2 superconformal transformations}
In this appendix we collect basic properties of 
$\cN$ = 1, 2 superconformal transformations
(see, for instance, refs. \cite{BK,osborn}
for more details). 
In $\cN$ = 1, 2  global superspace 
${\bf R}^{4|4\cN}$ 
parametrised by 
$z^A = (x^a, \q^\a_i, {\bar \q}^i_\ad) $,
infinitesimal superconformal transformations 
\be
z^A \quad \longrightarrow \quad z^A + \x^A
\ee
are generated by 
superconformal Killing vectors 
\be
\x = {\overline \x} = \x^A \,D_A =
\x^a (z) \pa_a + \x^\a_i (z)D^i_\a
+ {\bar \x}_\ad^i (z) {\bar D}^\ad_i
\ee   
defined to satisfy 
\be
[\x \;,\; D^i_\a ] \; \propto \; D^j_\b \;.
\ee   
${}$From here one gets
\be
\x^\a_i = -\frac{\rm i}{8} {\bar D}_{\bd i} \x^{\bd \a}\;, \qquad
{\bar D}_{\bd j} \x^\a_i = 0
\label{spinsc}
\ee
while the vector parameters satisfy the equation
\be
D^i_{(\a} \x_{\b )\bd} = {\bar D}_{i(\ad} \x_{\b \bd )}=0
\label{msc}
\ee
implying, in turn, the conformal Killing equation
\be
\pa_a \x_b + \pa_b \x_a = \hf\, \eta_{ab}\, \pa_c \x^c\;.
\ee 
${}$From eqs. (\ref{spinsc}) and (\ref{msc}) one gets
\be
[\x \;,\; D^i_\a ] = - (D^i_\a \x^\b_j) D^j_\b
= \hat{\o}_\a{}^\b  D^i_\b - \frac{1}{\cN}
\Big( (\cN-2) \s + 2 {\bar \s}  \Big) D^i_\a
-{\rm i} \hat{\L}_j{}^i \; D^j_\a \;.
\ee
Here the parameters of `local' Lorentz $\hat{\o}$ and
scale--chiral $\s$ transformations are
\be
\hat{\o}_{\a \b}(z) = -\frac{1}{\cN}\;D^i_{(\a} \x_{\b)i}\;,
\qquad \s (z) = \frac{1}{\cN (\cN - 4)}
\left( \hf (\cN-2) D^i_\a \x^\a_i - 
{\bar D}^\ad_i {\bar \x}_{\ad}^{ i} \right)
\label{lor,weyl}
\ee
and turn out to be chiral
\be
{\bar D}_{\ad i}\, \hat{\o}_{\a \b}~=~ 0\;,
\qquad {\bar D}_{\ad i}\, \s ~=~0\;.
\ee
The parameters $\hat{\L}_j{}^i$ 
\be
\hat{\L}_j{}^i (z) = -\frac{1}{32}\left(
[D^i_\a\;,{\bar D}_{\ad j}] - \frac{1}{\cN}
\d_j{}^i  [D^k_\a\;,{\bar D}_{\ad k}] \right)\x^{\ad \a}~, \qquad
\hat{\L}^\dag =  \hat{\L}~, \qquad  {\rm tr}\; \hat{\L} = 0
\ee
appear only in the $\cN=2$ case and 
correspond to `local' $SU(2)$ transformations.
One can readily check the identities 
\bea
D^k_\a \hat{\L}_j{}^i &=& 2{\rm i} \left( \d^k_j D^i_\a 
-\frac{1}{\cN} \d^i_j D^k_\a  \right) \s~, \non \\
D^i_\a \hat{\o}_{\b \g} &=& 2 \ve_{\a(\b}\, D^i_{\g)}\, \s~,
\eea
along with 
\be
D^i_\a  D^j_\b \,\s~ =~0~.
\ee

%\end{appendix}


\begin{thebibliography}{99}
\bi{DS}
M.~Dine and N.~Seiberg,
%``Comments on higher derivative operators in some SUSY field theories,''
Phys.\ Lett.\ {\bf B409} (1997) 239,
hep-th/9705057.

\bi{DG} M.~Dine and J.~Gray,
{\it Non-renormalization theorems for operators with 
arbitrary numbers of  derivatives in N = 4 Yang-Mills theory},
hep-th/9909020.

%\bi{PS} Paban Sethi  9806028



\bi{TA}
W.I.~Taylor,
{\it Lectures on D-branes, gauge theory and M(atrices)},
hep-th/9801182.
%%CITATION = HEP-TH 9801182;%%

\bi{CT} 
%Chepelev and Tseytlin 9709087
I.~Chepelev and A.A.~Tseytlin,
%``Long-distance interactions of branes: Correspondence 
%between  supergravity and super Yang-Mills descriptions,''
Nucl.\ Phys.\ {\bf B515} (1998) 73,
hep-th/9709087.

\bi{MM} 
J.M.~Maldacena,
%``Branes probing black holes,''
Nucl.\ Phys.\ Proc.\ Suppl.\ {\bf 68}, 17 (1998), 
hep-th/9709099.
%%CITATION = NUPHZ,68,17;%%



\bi{TR}  
%Tseytlin, 9908105
A.A.~Tseytlin,
{\it Born-Infeld action, supersymmetry and string theory},
hep-th/9908105.

\bi{BBPT} 
%Becker, Becker  Polchinski and Tseytlin
K.~Becker, M.~Becker, J.~Polchinski and A.~Tseytlin,
%``Higher order graviton scattering in M(atrix) theory,''
Phys.\ Rev.\ {\bf D56} (1997) 3174,
hep-th/9706072.

\bi{FT} 
%Fradkin and Tseytlin, 83
E.S.~Fradkin and A.A.~Tseytlin,
%``Quantum Properties Of Higher Dimensional 
%And Dimensionally Reduced Supersymmetric Theories,''
Nucl.\ Phys.\ {\bf B227} (1983) 252;
Phys.\ Lett.\ {\bf B123} (1983) 231.


\bi{MAL}   
%Maldacena  9711200
J.~Maldacena,
%``The large-N limit of superconformal field theories and supergravity,''
Adv.\ Theor.\ Math.\ Phys.\ {\bf 2} (1998) 231,
hep-th/9711200.

\bi{KAL} 
%Claus Kallosh et al 9801206
P.~Claus, R.~Kallosh, J.~Kumar, P.~Townsend and A.~Van Proeyen,
%``Conformal theory of M2, D3, M5 and D1+D5 branes,''
JHEP {\bf 06} (1998) 004,
hep-th/9801206.

\bi{MT}  
%Metsaev and Tseytlin 9806095
R.R.~Metsaev and A.A.~Tseytlin,
%``Supersymmetric D3 brane action in AdS(5) x S**5,''
Phys.\ Lett.\ {\bf B436} (1998) 281,
hep-th/9806095

\bi{gsw} R. Grimm, M. Sohnius and J. Wess,
Nucl. Phys. {\bf B133} (1978) 275.  

\bibitem{BK} I.L. Buchbinder and S.M. Kuzenko,
{\it Ideas and Methods of Supersymmetry and 
Supergravity} (IOP Publishing Ltd., Bristol, 1995, 1998). 

\bibitem{osborn} H. Osborn, Ann. Phys. {\bf 272} (1999)
243, hep-th/9808041;
%\bibitem{park2} 
J-H. Park, {\it Superconformal symmetry and
correlation functions}, hep-th/9903230;  
%\bibitem{kt} 
S.M. Kuzenko and S. Theisen, {\it Correlation 
functions of conserved currents in \mbox{$\cN=2$} superconformal theory},
hep-th/9907107.

\bibitem{gkpr} F. Gonzalez-Rey, B. Kulik, I.Y. Park and
M. Ro\v{c}ek, Nucl. Phys. {\bf B544} (1999) 218, 
hep-th/9810152.

\bibitem{bbk}
E.I. Buchbinder, I.L. Buchbinder and S.M. Kuzenko,
Phys.\ Lett.\ {\bf B446} (1999) 216,
hep-th/9810239.

\bibitem{lvu}
D.A. Lowe and R. von Unge,
JHEP {\bf 11} (1998) 014,
hep-th/9811017.

\bi{CTT}
%Chepelev and Tseytlin  hep-th/9705120
I.~Chepelev and A.A.~Tseytlin,
%``Interactions of type IIB D-branes from the D-instanton
%matrix model,''
Nucl.\ Phys.\ {\bf B511} (1998) 629,
hep-th/9705120.

\bibitem{bbko} I.L. Buchbinder, E.I. Buchbinder,
S.M. Kuzenko and B. A. Ovrut, Phys. Lett. {\bf B417} (1998) 61,
hep-th/9704214.

\bibitem{gikos} A. Galperin, E. Ivanov, S. Kalitzin, V. Ogievetsky 
and E. Sokatchev, Class. Quantum Grav. {\bf 1} (1984) 469.

\bibitem{antal}
A.~Jevicki, Y.~Kazama and T.~Yoneya,
%``Quantum metamorphosis of conformal transformation in
%D3-brane  Yang-Mills theory,''
Phys.\ Rev.\ Lett.\ {\bf 81}, 5072 (1998), 
hep-th/9808039.
%%CITATION = PRLTA,81,5072;%%

\bibitem{bk1}
I.L. Buchbinder and  S.M. Kuzenko, 
 Mod. Phys. Lett. 
{\bf A13}, 1623 (1998), hep-th/9804168. 

\bibitem{opb}
T. Ohrndorf, Phys. Lett. {\bf B176} (1986) 421; 
N.G. Pletnev and A.T. Banin, Phys. Rev. {\bf D60} 
(1999) 105017,
hep-th/9811031.

\bibitem{schwinger} J.S. Schwinger, Phys. Rev. 
{\bf 82} (1951) 664.

\bibitem{gr}
F.~Gonzalez-Rey and M.~Ro\v{c}ek,
Phys.\ Lett.\ {\bf B434} (1998) 303,
hep-th/9804010.

\bibitem{bi} S. Cecotti and S. Ferrara, Phys. Lett.
{\bf B187} (1987) 335; 
J.~Bagger and A.~Galperin,
%``A new Goldstone multiplet for partially broken supersymmetry,''
Phys.\ Rev.\ {\bf D55} (1997) 1091,
hep-th/9608177;
M.~Ro\v{c}ek and A.A.~Tseytlin,
%``Partial breaking of global D = 4 supersymmetry, 
%constrained  superfields, and 3-brane actions,''
Phys.\ Rev.\ {\bf D59} (1999) 106001,
hep-th/9811232.

\bibitem{hsw} P.S. Howe, K.S. Stelle and P.C. West,
Phys. Lett. {\bf 124B} (1983) 55.

\bibitem{sqed} 
K. Shizuya and Y. Yasui, Phys. Rev. {\bf D29} (1984) 1160;
I.L. Buchbinder and S.M. Kuzenko,
Sov. Phys. J. (Izv. VUZov, Fiz.) {\bf 28} (1985) 64.

\bibitem{mag}
I.N. McArthur and T.D. Gargett, Nucl. Phys.
{\bf B494} (1997) 525, hep-th/9705200.


\bi{bky}I.L.~Buchbinder, S.~Kuzenko and Z.~Yarevskaya,
%``Supersymmetric effective potential: Superfield approach,''
Nucl.\ Phys.\ {\bf B411} (1994) 665.

\bi{bbiko}
I.L.~Buchbinder, E.I.~Buchbinder, E.A.~Ivanov, S.M.~Kuzenko and B.A.~Ovrut,
%``Effective action of the N = 2 Maxwell multiplet in harmonic superspace,''
Phys.\ Lett.\ {\bf B412} (1997) 309,
hep-th/9703147.
%%CITATION = PHLTA,B412,309;%%


\end{thebibliography}
\end{document}